\documentclass[10pt,a4paper,onecolumn,numbers]{elsarticle}
\usepackage[latin1]{inputenc}
\usepackage{amsmath}
\usepackage{amsfonts}
\usepackage{amssymb}
\usepackage{graphicx}
\usepackage{fancyhdr}
\usepackage{times}
\usepackage{natbib}

\begin{document}
\begin{frontmatter}

\title{Reconstructing energy and X$_{\mathrm{max}}$ of cosmic ray air showers using the radio lateral distribution measured with LOPES}


\author[1]{W.D.~Apel}
\author[2,14]{J.C.~Arteaga}
\author[3]{L.~B\"ahren}
\author[1]{K.~Bekk}
\author[4]{M.~Bertaina}
\author[5]{P.L.~Biermann}
\author[1,2]{J.~Bl\"umer}
\author[1]{H.~Bozdog}
\author[6]{I.M.~Brancus}
\author[4]{A.~Chiavassa}
\author[1]{K.~Daumiller}
\author[2,15]{V.~de~Souza}
\author[4]{F.~Di~Pierro}
\author[1]{P.~Doll}
\author[1]{R.~Engel}
\author[3,9,5]{H.~Falcke}
\author[2]{B.~Fuchs}
\author[10]{D.~Fuhrmann}
\author[11]{H.~Gemmeke}
\author[7]{C.~Grupen}
\author[1]{A.~Haungs}
\author[1]{D.~Heck}
\author[3]{J.R.~H\"orandel}
\author[5]{A.~Horneffer}
\author[2]{D.~Huber}
\author[1]{T.~Huege}
\author[1,16]{P.G.~Isar}
\author[10]{K.H.~Kampert}
\author[2]{D.~Kang}
\author[11]{O.~Kr\"omer}
\author[3]{J.~Kuijpers}
\author[2]{K.~Link}
\author[12]{P.~{\L}uczak}
\author[2]{M.~Ludwig}
\author[1]{H.J.~Mathes}
\author[2]{M.~Melissas}
\author[8]{C.~Morello}
\author[2]{J.~Oehlschl\"ager}
\author[2]{N.~Palmieri\corref{cor}}
\ead{Nunzia.Palmieri@kit.edu}
\author[1]{T.~Pierog}
\author[10]{J.~Rautenberg}
\author[1]{H.~Rebel}
\author[1]{M.~Roth}
\author[11]{C.~R\"uhle}
\author[6]{A.~Saftoiu}
\author[1]{H.~Schieler}
\author[11]{A.~Schmid}
\author[1]{F.G.~Schr\"oder}
\author[13]{O.~Sima}
\author[6]{G.~Toma}
\author[8]{G.C.~Trinchero}
\author[1]{A.~Weindl}
\author[1]{J.~Wochele}
\author[1]{M.~Wommer}
\author[12]{J.~Zabierowski}
\author[5]{J.A.~Zensus}

\address[1]{Institut f\"ur Kernphysik, Karlsruhe Institute of Technology (KIT), Germany}
\address[2]{Institut f\"ur Experimentelle Kernphysik, KIT - Karlsruher Institut f\"ur Technologie, Germany}
\address[3]{Radboud University Nijmegen, Department of Astrophysics, The Netherlands}
\address[4]{Dipartimento di Fisica Generale dell' Universit\`a di Torino, Italy}
\address[5]{Max-Planck-Institut f\"ur Radioastronomie Bonn, Germany}
\address[6]{National Institute of Physics and Nuclear Engineering, Bucharest, Romania}
\address[7]{Fachbereich Physik, Universit\"at Siegen, Germany}
\address[8]{Istituto di Fisica dello Spazio Interplanetario, INAF Torino, Italy}
\address[9]{ASTRON, Dwingeloo, The Netherlands}
\address[10]{Fachbereich Physik, Universit\"at Wuppertal, Germany}
\address[11]{Institut f\"ur Prozessdatenverarbeitung und Elektronik, KIT - Karlsruher Institut f\"ur Technologie, Germany}
\address[12]{Soltan Institute for Nuclear Studies, Lodz, Poland}
\address[13]{Department of Physics, University of Bucharest, Bucharest, Romania}
\address[14]{now at: Univ Michoacana, Morelia, Mexico}
\address[15]{now at: Univ S$\tilde{a}$o Paulo, Inst. de F\'{\i}sica de S\~ao Carlos, Brasil}
\address[16]{now at: Inst. Space Sciences, Bucharest, Romania }

\begin{abstract}
%In the previous decades, remarkable progress has been made in the detection of electromagnetic emission from cosmic ray air showers. 
The LOPES experiment, a digital radio interferometer located at KIT (Karlsruhe Institute of Technology), obtained remarkable results for the detection of radio emission from extensive air showers at MHz frequencies.\\
%Aiming to become competitive with the well-established investigation methods, radio detection has the main purpose of retrieving the complete information from a high-energy cosmic ray, e.g. arrival direction, energy and type of the primary particle.\\
Features of the radio lateral distribution function (LDF) measured by LOPES are explored in this work for a precise reconstruction of two fundamental air shower parameters: the primary energy and the shower X$_{\mathrm{max}}$.\\
The method presented here has been developed on (REAS3-)simulations, and is applied to LOPES measurements. Despite the high human-made noise at the LOPES site, it is possible to reconstruct both the energy and X$_{\mathrm{max}}$ for individual events. On the one hand, the energy resolution is promising and comparable to the one of the co-located KASCADE-Grande experiment. On the other hand, X$_{\mathrm{max}}$ values are reconstructed with the LOPES measurements with a resolution of 90\,g/cm$^{2}$. A precision on X$_{\mathrm{max}}$ better than 30\,g/cm$^{2}$ is predicted and achievable in a region with a lower human-made noise level.
\end{abstract}
\end{frontmatter}
%\begin{document}
%\maketitle

\section{Introduction}
Knowing the type of cosmic rays which interact in the atmosphere with an energy larger than 10$^{14}$\,eV still remains a fundamental goal in cosmic ray physics. A precise knowledge of the mass composition for the complete energy spectrum will help in distinguishing between several models for cosmic ray origin and propagation.\\
The detection of radio emission from cosmic ray air showers in the MHz regime as well as the understanding of its emission mechanisms made impressive progress in the recent years.\\ 
The LOPES experiment \cite{lopes} is one of the pioneering in radio detection and still supplies us with remarkable results \cite{frank3}. One of its main advantage is the co-location with the particle detector KASCADE-Grande \cite{kg} at KIT, Germany. \\
Recently, two independent methods have been separately investigated and tested on the LOPES data in order to extract X$_{\mathrm{max}}$, i.e. atmospheric depth of the shower maximum, from radio measurements and thus, indirectly, to achieve information on the type of primary cosmic rays. One method considers the shape of the radio shower front \cite{frank2,frank3}, the other method is presented in the following (slope method) and uses the slope of the radio lateral distribution function (LDF), i.e. the radio amplitudes at several distances from the shower axis \footnote{Hence the name ``slope method"}. \\
The correlation between the slope of the radio LDF and the shower maximum depth has been predicted since years from simulations \cite{REAS3,MGMRc}. This dependence can easily be referred to a geometrical effect: iron nuclei interact earlier in the atmosphere and develop faster compared to proton primaries; the radio source for an iron event is, thus, further away from an observer at ground, and, as a consequence, the lateral distribution function slope is flatter compared to proton events.\\
Taking the analysis in \cite{REAS2c} as guideline, a simulation-based method (slope method) which uses the LDF slope to extract two fundamental shower parameters (primary energy and X$_{\mathrm{max}}$) is developed and applied to LOPES data. The main results are presented in the following. 

\section{LOPES events}
The LOPES experiment \cite{lopes} is a digital interferometric radio antenna array placed at KIT, Germany. The events selected for the slope method analysis have been measured with the LOPES30 and LOPESpol setups \cite{timArena}. The first consisted of 30 calibrated dipole antennas, all oriented in the east-west direction, while the second used only 15 antennas aligned in the east-west direction. Due to a higher statistics of events expected in the east-west aligned antennas compared to the north-south direction \cite{geomagnEffect}, only the analysis on east-west detected events is shown. All the antennas operate in the effective frequency range of 43-74\,MHz.\\
LOPES profits from the reconstruction of air shower parameters, such as primary energy, shower core and incoming direction, from the particle detector array KASCADE-Grande \cite{kg}.\\
In the selected events the primary energy is around $\sim 10^{17}$\,eV and the zenith angle is less then 40\,deg. The core position of the shower is required to be at a distance of at most 90\,m from the center of the LOPES array, in order to avoid events with amplitudes in the tail of the LDF, thus affected by larger fluctuations. High signal-to-noise and high coherency for the radio signal in the antennas is required as well. Further qualitative cuts demand a good fit for the lateral distribution function, i.e. small $\chi^{2}$.
Over 200 events are selected in this way. \\
The radio lateral distribution for each individual event is fit with an exponential function, which is considered a good approximation for the radio lateral behaviour at the distances probed by LOPES \cite{Nunzia_thesis}.\\
%Thanks to the precise amplitude calibration of each individual LOPES antennas and 
Improvements are constantly made in modeling the radio emission from extensive air showers. The results shown in the following are based on REAS3 simulations (REAS3.11 and CoREAS are the most recent versions \cite{coreas,coreas2}), which already showed a good agreement with the LOPES measured LDFs for almost all the events \cite{reas3}.\\

\section{The slope method}
\begin{figure}[h!]
  \vspace{5mm}
  \centering
  \includegraphics[width=1\textwidth]{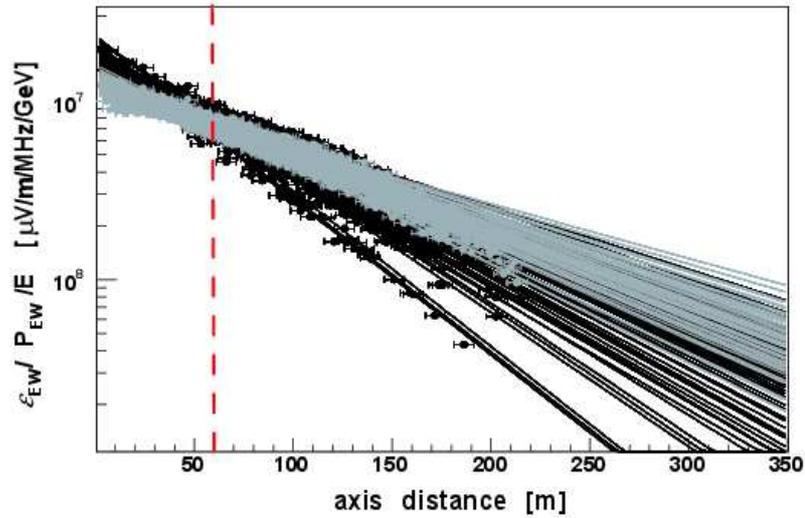}
  \caption{Normalized REAS3 lateral distribution function for the events with zenith angle smaller than 20\,deg., simulated once as proton (black), once as iron (gray). The points are fitted with an exponential function.}
   \label{LDF}
 %    \includegraphics[width=3.in, height=2in]{rms.eps}
 %  \caption{....}
 % \label{rms}
\end{figure} 
The slope method has been developed with REAS3 simulations, and it is then applied to LOPES data. For each LOPES measured event, one proton-generated air shower and one iron-generated air shower is produced with CORSIKA \cite{corsika} (QGSJetII is used as hadronic interaction model). Afterwards, REAS3 \cite{reas3} generates the radio emission for the CORSIKA simulated air showers.\\
For the purpose of this analysis, the simulations are created in order to reproduce a realistic case: 
on the one hand, information about the LOPES selected events, such as core position, primary energy, the number of muons (N$\mu$), and incoming direction reconstructed by KASCADE(-Grande) are used as input parameters for the CORSIKA showers \cite{reas3}.\\
 On the other hand, shower-to-shower fluctuation are included, since not a typical shower, i.e. with a typical X$_{\mathrm{max}}$, is selected. In order to represent at best the recorded event, a further step is made in the pre-selection of the CORSIKA showers: the N$\mu$ measured by KASCADE(-Grande) is used as discriminator for this purpose. 200 CONEX showers for proton and 100 for iron are simulated with QGSJetII and UrQMD respectively for high and low energy interaction\cite{corsika,reas3}. Among all, the CONEX shower which can best reproduce the measured N$\mu$ is chosen. In this way a specific shower similar to what has been detected by KASCADE is used.\\
Moreover, the observer positions for the REAS3-simulated LDF represent the real arrangement of the LOPES antennas in the field with respect to the core of the shower.\\ 

As for the LOPES events, an exponential fit is applied to each REAS3 lateral distribution functions.\\
The inclination of the air shower, as well as the type of the primary particle, considerably affect the slope of the radio LDF. Again this is related to a geometrical effect: for larger zenith angle, the radio source is further away from the observer at ground, thus a flatter slope for the LDFs is expected compared to vertical air showers. Thereby, several zenith angle bins are separately considered, and only the plots for the first zenith bin (up to 20\,deg.) are shown here.\\
The slope method aims to compare LDFs of events (in the same zenith bin) with different primary energies and arrival directions, thus normalizations for the radio amplitudes are required. The primary energy reconstructed by KASCADE(-Grande) is used for the energy normalization, while the corrections for the arrival direction involve the Lorentz force vector (precisely the east-west component of the vector Pew). This last implies the reasonable assumption that the predominant contribution to the radio emission in air showers has geomagnetic origin \cite{geomagnEffect,Nunzia_thesis}.\\ 
The LDFs for 54 events (first zenith bin) are shown in Fig.~\ref{LDF} and the difference between irons and protons is clearly visible by eye.\\
   
\begin{figure}[ht!]
  \vspace{5mm}
  \centering
  %\begin{minipage}[t]{width=3.in}
   \includegraphics[width=1\textwidth]{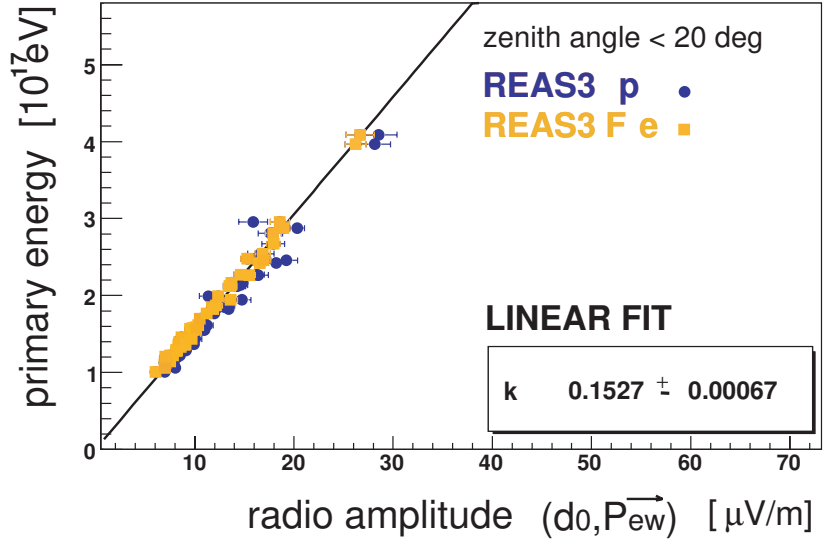}
  \caption{The correlation between the reconstructed primary energy and the REAS3 radio pulse in the flat region (d$_{0}$) is fit with a linear function, with k$_{l}$ the free parameter. The radio amplitude, in the east-west direction, is normalized by the single component of the Lorentz vector. The RMS-spread is of $\sim$6$\%$.}
   \label{REAS_energy_lin}  
  %\end{minipage}
\end{figure} 

\subsection{Primary energy correlation}
A specific region (d$_{0}$), so called flat region,  where all the LDFs profiles intersect and where the radio amplitude does not carry any information about the primary type is expected \cite{REAS2c}. This region is identified by looking at the RMS spread of the LDF fits at 12 distances from the shower core and picking the smallest RMS value. For the LOPES events d$_{\mathrm{0}}$ happens in the distance range between 70 and 90\,m from the shower axis \footnote{in shower plane coordinate system} depending on the zenith angle bin, and it is marked by the dashed line in Fig.~\ref{LDF} (60\,m).\\
From previous analyses \cite{REAS2c}, a direct correlation is expected between the radio amplitude in d$_{\mathrm{0}}$ - normalized for the arrival direction with Pew (as explained above)- and the energy, in particular with the fraction of the energy from the electromagnetic component of the air shower. This is due to the intrinsic nature of the radio emission.\\
For the LOPES experiment, only the total primary energy reconstructed by KASCADE(-Grande) is available (Fig.~\ref{REAS_energy_lin}). A linear fit is used for this correlation. The spread around the fit is an indicator of the precision one acquires in determining the energy with this method. For this LOPES selection, an RMS spread of maximum 8$\%$ is found for the complete LOPES selection ($\sim$6$\%$ in Fig.~\ref{REAS_energy_lin} for zenith angle $<$20\,deg.) for the total primary energy correlation. Much higher precision - RMS spread $<4\%$ - is predicted for the electromagnetic energy (not shown here) \cite{Nunzia_thesis}.     
\begin{figure}[h!]
  \vspace{5mm}
  \centering
  \includegraphics[width=1\textwidth]{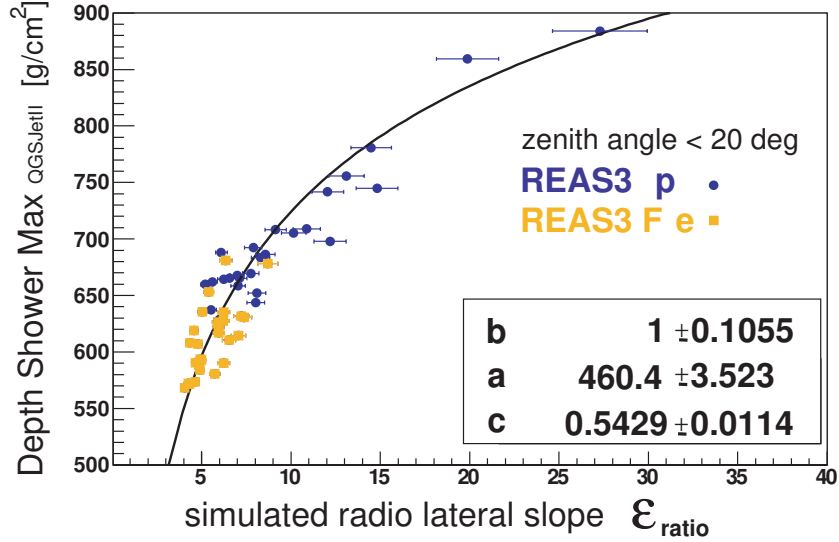}%ICRC2011/submitted/icrc0309/icrc0309_fig03.eps}
  \caption{Correlation between the true CORSIKA X$_{\mathrm{max}}$ and the radio simulated LDF slope, for proton (blue) and iron (red) simulated primaries. The RMS-spread is of $\sim$29 gcm$^{-2}$.}
   \label{REAS_Xmax}
\end{figure}

\subsection{X$_{\mathrm{max}}$ correlation}
The slope of the radio LDFs is sensitive to the depth of the shower maximum (X$_{\mathrm{max}}$). Independently of the function used to fit the LDF, the ratio ($\epsilon_{\mathrm{ratio}}$) of the radio amplitudes at two different distances is an indicator of the LDF slope. In this analysis the amplitudes at d$_{\mathrm{0}}$ and at d$_{\mathrm{0}}$+170m are considered.\\
The function Eq.~\ref{xm_eq} \cite{REAS2c} is used to fit the correlation
\begin{equation}
X_{max} = a \left[\mathrm{ln} \left(b \epsilon_{\mathrm{ratio}}\right)\right]^{c}
\label{xm_eq}
\end{equation}
between REAS3 $\epsilon_{\mathrm{ratio}}$ and X$_{\mathrm{max}}$ from CORSIKA simulations, separately for each zenith angle bin (Fig.~\ref{REAS_Xmax}). The RMS spread is an indicator of the uncertainty on X$_{\mathrm{max}}$ reconstruction with the slope method; for the complete zenith angle range an uncertainty of 20-40\,g/cm$^{2}$ is predicted, with the larger values due to the larger zenith angles.\\
The values for the three fitting parameters (\textit{a, b} and \textit{c}) will be used in the following to reconstruct X$_{\mathrm{max}}$ with the LOPES measurements.  
\newpage

\begin{figure}[h!]
  \vspace{5mm}
  \centering
  \includegraphics[width=1\textwidth]{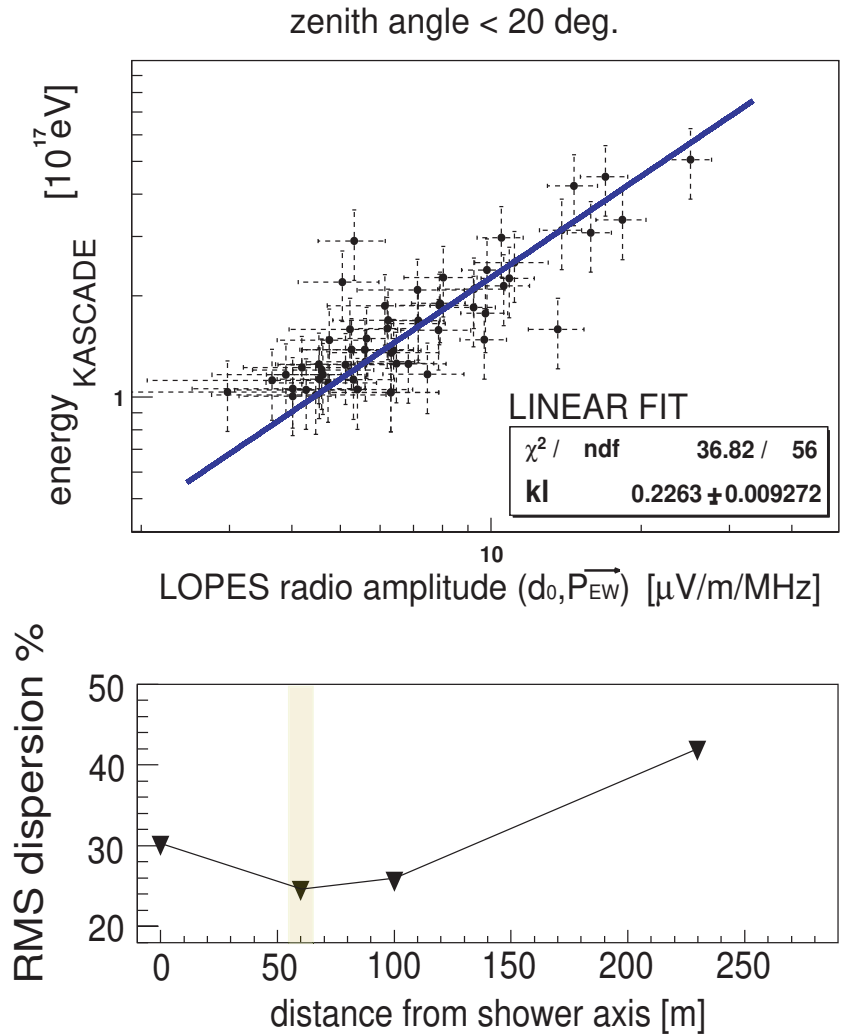}
  \caption{TOP: Linear correlation of the KASCADE-Grande reconstructed primary energy and the LOPES measured radio pulse in d$_{0}$, normalized for Pew. The RMS-spread if of $\sim23\%$ BOTTOM: Comparison of the RMS-spread for the linear energy fit computed at 4 distances from the shower axis, for zenith angle $<$20\,deg. In d$_{0}$ the RMS-spread has the smallest value. The LOPES measurements confirm the distinctive feature of d$_{0}$. }
   \label{LOPES_en}
\end{figure} 

\section{LOPES measurements}
One of the main results already achieved with the LOPES experiment is the clear dependence of the recorded radio pulse (CC-beam amplitude) with the energy of the primary cosmic ray \cite{geomagnEffect}.
In the following another approach for the primary energy identification by using the radio signal at a well specific distance from the shower axis is presented.\\
The same distances d$_{\mathrm{0}}$, previously identified for each zenith angle bin with the REAS3 simulations, are used for the LOPES measurements as well. For each individual event, the LOPES $\epsilon_{\mathrm{d_{0}}, Pew}$ is the value of the fit of the LOPES lateral distribution function - corrected for the arrival direction - at the predicted distance d$_{\mathrm{0}}$. As for the REAS3 simulations, the linear correlation with the primary energy reconstructed by KASCADE(-Grande) is shown (Fig.~\ref{LOPES_en} - top side). The uncertainty on the energy reconstruction is again identified by the RMS spread, and it is of about 20$\%$ averaging over the complete selection, i.e. zenith angle between 0 and 40\,deg. This value is comparable with the statistical uncertainty of KASCADE \cite{kg}.\\
As a cross-check of the real existence in the measurements of this preferable distance for the energy reconstruction, the energy correlation is investigated considering the radio amplitudes at three further distances from the shower axis: 0, 100, d$_{\mathrm{0}}$+170\,m (Fig.~\ref{LOPES_en} - bottom side). The RMS spread to the linear fit results to be lowest when the radio amplitude is taken at d$_{\mathrm{0}}$, confirming the expectations.

\begin{figure}
  \vspace{5mm}
  \centering
  \includegraphics[width=1\textwidth]{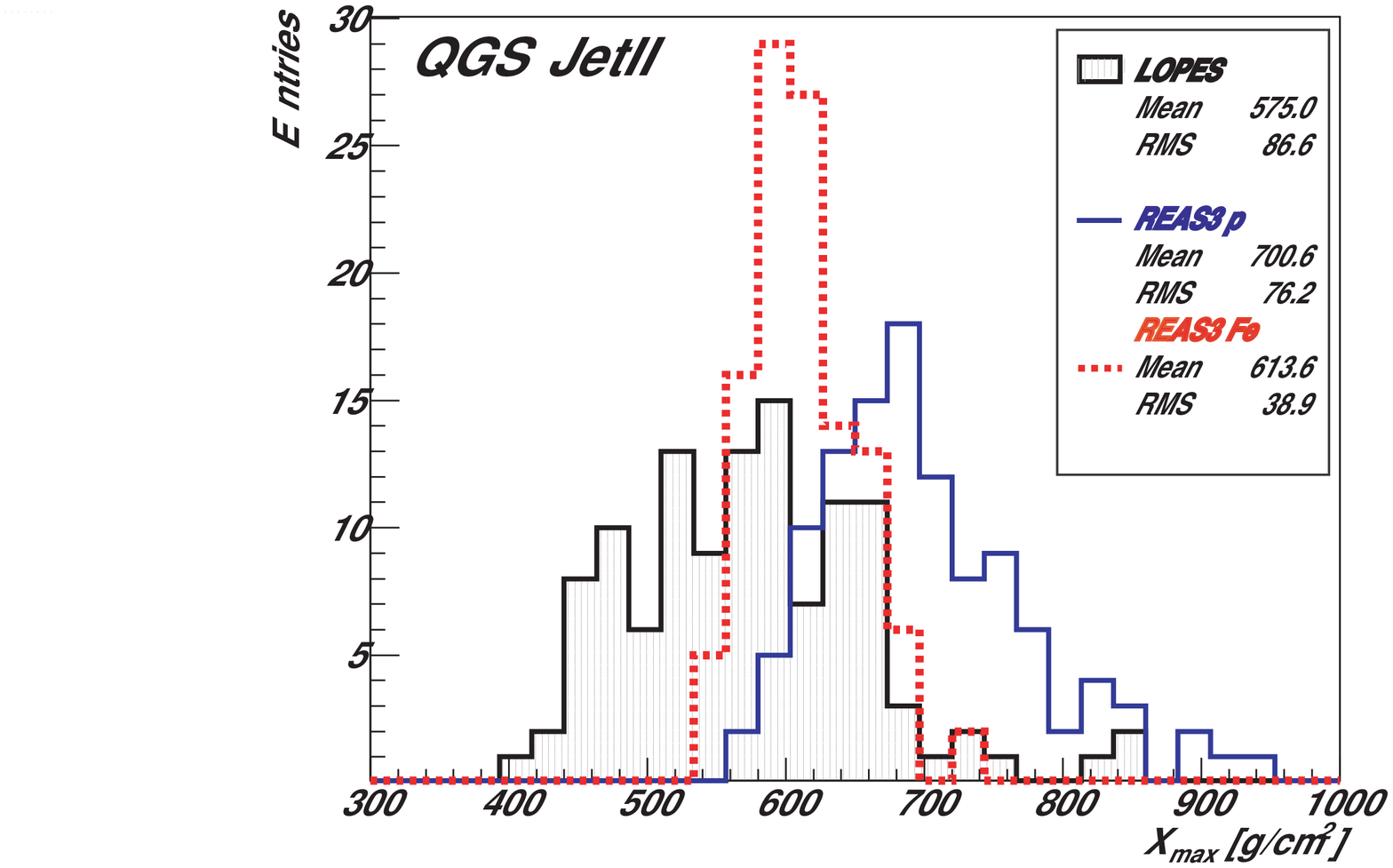}%ICRC2011/submitted/icrc0309/icrc0309_fig05.eps}
  \caption{X$_{\mathrm{max}}$ reconstructed with the slope method for the LOPES measurements (black) and for REAS3 simulations (proton (blue) and iron (red)), for the complete zenith angle range- 0-40\,deg. }
   \label{LOPES_xm}
\end{figure} 

The X$_{\mathrm{max, LOPES}}$ are reconstructed for the complete LOPES selection using the Eq.~\ref{xm_eq} and the \textit{a, b} and \textit{c} identified with the REAS3 simulations. In Figure~\ref{LOPES_xm}, X$_{\mathrm{max, LOPES}}$ of 600~$\pm$~90\,g/cm$^{2}$, i.e. mean and standard deviation values. %In the real case of a mixed composition selection, the statistical uncertainty of the LOPES X$_{\mathrm{max, LOPES}}$ is comparable with the REAS3 prediction. Moreover, 
The LOPES reconstructed X$_{\mathrm{max}}$ values are almost compatible with the expectations from the cosmic ray nuclei. However, they are shifted to X$_{\mathrm{max}}$ smaller than the REAS3 iron-like predictions. This shift is surely influenced by the still existing systematic divergence between the slope of the simulated (REAS3) and the measured (LOPES) lateral distribution functions. The most recent REAS3.11 and CoREAS simulations have already shown a better agreements with the measurements, therefore improvements are expected by the upcoming application of the slope method to the newest simulations.
%\subsection{Primary energy estimation}
%\subsection{X$_{\mathrm{max}}$ reconstruction}
\section{conclusion}
The slope method has been successfully applied to REAS3 simulations and measurements of LOPES data.\\
This method reveals itself to be a powerful tool for both energy and mass composition investigations with the radio data.\\
A well specific distance from the shower axis is confirmed even in the LOPES measurements to be the best place for the primary energy reconstruction. A precision of almost 20$\%$ in energy reconstruction is found for the LOPES data and it is comparable to the statistical uncertainty of the KASCADE experiment.\\
X$_{\mathrm{max}}$ values obtained for the LOPES measured data are almost comparable with expectations. Systematic divergences are also discussed and improvements are foreseen with REAS3.11 and CoREAS simulations.\\ An upper-limit precision of 90\,g/cm$^{2}$ is found with the LOPES measurements, being the highest precision on X$_{\mathrm{max}}$ sensitivity to date achievable with the radio data.\\
Despite the high environmental noise which limits the performance of the LOPES experiment, important results are found. \\
Higher precision in both energy and X$_{\mathrm{max}}$ reconstruction predicted by the slope method may be achieved with a negligible noise background. Especially in the case of AERA (Auger Engineering Radio Array \cite{aera}), located at the Pierre Auger Observatory, which could cross-check the reconstructed X$_{\mathrm{max}}$ with the experimental values obtained by the Fluorescence Detector (FD). 

\hspace{2cm}

%\begin{theacknowledgments}
\textbf{Acknowledgments} LOPES and KASCADE-Grande have been supported by the German Federal Ministry of Education and Research. KASCADE-Grande is partly supported by the MIUR and INAF of Italy, the Polish Ministry of Science and Higher Education and by the Romanian Authority for Scientific Research UEFISCDI (PNII-IDEI grant 271/2011). This research has been supported by grant number VH-NG-413 of the Helmholtz Association.
%\end{theacknowledgments}

\bibliographystyle{aipproc}
%\bibliography{arena2012}

\vspace{\baselineskip}
\

\end{document}